# Safety Requirement Specifications for Connected Vehicles

Madhusudan Singh,

Yonsei Institute of Convergence Technology,

Yonsei University, Song-Do, Yeonsu-Gu, Korea

msingh@yonsei.ac.kr

Shiho Kim,

Yonsei Institute of Convergence Technology,

Yonsei University, Song-Do, Yeonsu-Gu, Korea

shiho@yonsei.ac.kr

## ABSTRACT

In the coming years, transportation system will be revamped in a manner that there will be more intelligent and autonomous vehicle phenomenon around us such as smart cars, auto driving system, etc. Some of automotive industries are already producing smart cars. However, the main concern of this paper is on the infrastructure for connected vehicles, which can support such intelligent transportation. Current transportation system lacks proper infrastructure to support connected vehicles. Hence, in this article, we have surveyed and analyzed the current transportation system in developed and developing countries. In contrast, we are going to introduce secure intelligent transportation (roadside) infrastructure that is user centric (Driver, Autonomous driver etc.) for connected vehicles. In this paper we present the basic requirements of safety engineering infrastructure of roadside infrastructure in ITS for connected vehicles. Connected vehicles has network infrastructure to communicate with vehicle-to-vehicle (V-to-V), vehicle-to-infrastructure (V-to-I), lane correction system, and traffic information system etc. The connected vehicle is a good model for learning demands of infrastructure for ITS process because the system having a lot of use-cases and we must understand relationship between public institutions, people, companies in order to proceed ITS System.

## 1. INTRODUCTION

Intelligent Transportation System (ITS) is the term of the automobile and connected vehicles that have intelligence ability including computing, measuring, situation judging and networking something. Therefore, ITS attracts the public attention by evolving from simple technology to containing convergence technologies where smart can judge it. The most interesting thing is that, the connected vehicles can judge their decision. ITS raise new business trend in all the world market place through uniting Information communication technology (ICT) and automobile that lead the hardware, software and manufacturing industries [1]. Nowadays, after making convention with automotive manufacturer like BMW, Audi, Honda, etc. and the huge IT Company like Google, Microsoft, Apple etc. released automobile operating system as-well-as automotive manufacturers constructed environment for making intelligent vehicle based on the advanced technologies like automatic parking system, lane- keeping system, conflict prevention system, and so on. With these developments, combined with the needs of consumers, it required establishing of transport infrastructure that is used for operating intelligent automotive [2-4].

After starting full-scale vehicle life in the 1970s, society has required the development of transport system for safety and efficiency. Automotive industry has obtained not only accelerating industrialization but also expanding of living area. However, behind the scenes, social costs were increased by traffic accidents. In fig.1 shows that the no. of accidents in major vehicles consuming countries [5]. To prevent these situations, we need a more advanced ITS system then the current scenario. Especially, we must accept to development of relevant market for economic benefit, after preoccupying market with high relevant technologies. So, government started to develop network system called Smart-Highway that have a seamless communication system on high speeds [6].

This paper presents the basic requirements of safety engineering infrastructure of roadside infrastructure in ITS for connected vehicles. Connected vehicles has network infrastructure to communicate with vehicle-to-vehicle (V-to- V), vehicle-to-infrastructure (V-to-I), lane correction system, and traffic information system etc. The connected vehicles have a good model for learning demands of infrastructure for ITS process because the system have a lot of use-cases and we must understand relationship between public institutions, people, companies in order to proceed ITS System.

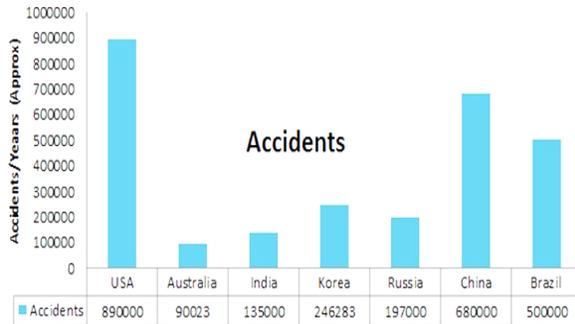

Fig.1. Number of accidents in major vehicles consuming countries.

This paper has organized into sections as follows: Section II provides problem in connected vehicles and their challenges in V-to-V communication system. Section III, provides the requirement specifications of safe and secure engineering design architecture. In the section IV, we discussed the novel safe and secure engineering model for connected vehicles. Finally, in the section V, we have concluded final remarks and impact of connected vehicles with the respect of social, technical and business.

## 2. PROBLEM STATEMENTS AND CHALLENGES

Current ITS technique optimize traffic management and highway capacity by introducing spatial-temporal distribution of traffic flow to provide information through various media such as VMS, broadcast, internet, and etc. after ITS collects real-time traffic information through detection devices and CCTV on the highway in 2,804km of 23 routes. However, republic of Korea has operated highway traffic management system, since public infrastructure investment planning group to the president introduced ITS in 1993 [7]. However, ITS infrastructure has to consider road, IT, and V-to-V communication such as shown in fig.2.

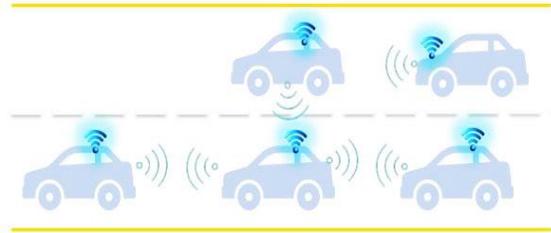

Fig.2. Vehicle-to-Vehicle communication system.

ITS of the Republic of Korea has automatic vehicle identification system, image detection system and real time tracking algorithm through traffic monitoring camera such as CCTV and traffic related operation center that provides traffic information service through wire and wireless communication by building comprehensive traffic management system that has safety and quickness [7]. These find violators by exchanging information with each other through optical communication and wireless network, and maintain traffic flow system by transmitting traffic flow data in real-time. In particular, ITS of the Republic of Korea improve traffic flow overhead that is produced by tollgate, and reach next generation ITS environment because of electronic toll collection system because Hi-pass payment system using DSRC was introduced to ITS of the Republic of Korea in 2007 [8].

### A. Current Infrastructure and Challenges in ITS

- **Information content:** There is no condition of information exchange type between road and vehicle Therefore we need to provide point-based traffic information (VMS-oriented) and indirect information such as ARS, Internet, etc.

- **Communications:** There is no demand control function for preventing excess capacity. Therefore, Lack of traffic flow distributed technology of whole road in case of emergency, Lack of real-time sensing function in case of emergency, Lack of estimated function of traffic, condition when delaying.

- **Security:** Restrictive crackdown of unit point about Violator's vehicles which are the main culprit of big accident passive management systems such as use of manpower when road maintenance management (safety issue)

- Others : Interchange (IC) system that abnormally add to the ETC, Difficulty of differentiated Services about the ETC vehicle, Limited payment structure (prepaid), uniform way to settle the charges on IC system, Lack of velocity-based and high-tech traffic condition model, However, there is no information service for road manager.

## 3. SPECIFICATIONS OF SAFETY ENGINEERING

The basic requirements of intelligent transportation world such as intelligent transportation system, Infrastructure, smart vehicles, connected vehicles requirements and their challenges are given in fig. 3.

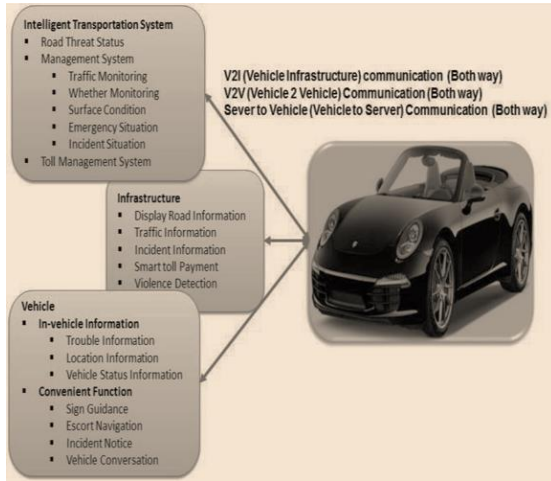

Fig. 3. Specification of connected vehicles.

However, ITS used to collect road information from various Road Infrastructures and analyses it to offered various traffic information and incident situation to Road Infrastructure and vehicles from the analysed data. Infrastructure information is required prevent dangerous situation by detecting cargo fall, vehicle stopped by accident, jaywalker detected and various incident situations with camera and Radar. By send collected vehicle speed and traffic information to the Center, It also can be offered to vehicles and displays that escort navigation service. Smart tolling systems charge the fee through Tag of vehicles and also can be charged differently by detecting class of vehicles. Vehicle can be prevented potential dangerous situation by using collected in-vehicle data. Display road status and sign information to in-vehicle terminal from Road Infrastructure.

A. *Technology requirements of Connected Vehicles in ITS*

1. ITS comprehensive system.
2. Smart information and traffic management system implemented direct information exchange environment between road and vehicle.
3. Continuous roadside wireless communication system for both Call & Response implementations.
- Collect and provide traffic information Seamless based on roadside network
- Configure a wireless multi-hop network for various telematics services, safety of drivers that drive at high speed
- Smart roadside communication infrastructure that can receive continuous information offering in the vehicle
- Smart vehicle terminal and content that provide Call & Response to each individual vehicle
- Smart business system based on Non-stop and a number of roads

## 4. ENGINEERING MODEL FOR SEAFTY

The overview infrastructure of a secure connected vehicles in ITS has shown in fig. 4, where the communication tower collect the data such as traffic data, weather data on highway, car data on highway based on data of mobile device that mounted in the car, radio tower send data to main data normalization and event analyzers system. After the analysis of data, the information forwarded to the information storage cloud to mobile device again. Then mobile device provides user interfaces to check easier with input value of the driver (user). However, Fig. 5 is context diagram that shows an overview infrastructure of the system that we proposed

The major requirements of the telematics based safe and connected vehicles system are required four keywords of system, which is as follows:

- Subject roadside: The subject matter of the information system.
- System roadside information: Charge, traffic jam and so on.
- Usage Driver: The environment within which the planned system will operate.
- Network System: Driver who use the highway, and stakeholder who related on the highway system.
- System intelligent roadside: What the system does within its operational environment, what information it contains and what function it performs.

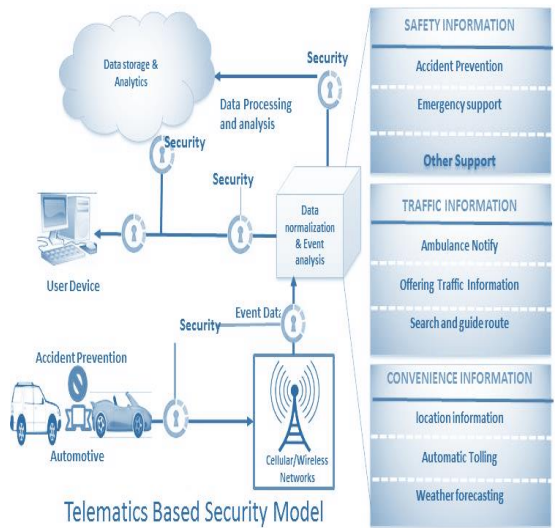

Fig. 4. Telematics based safe connected vehicle

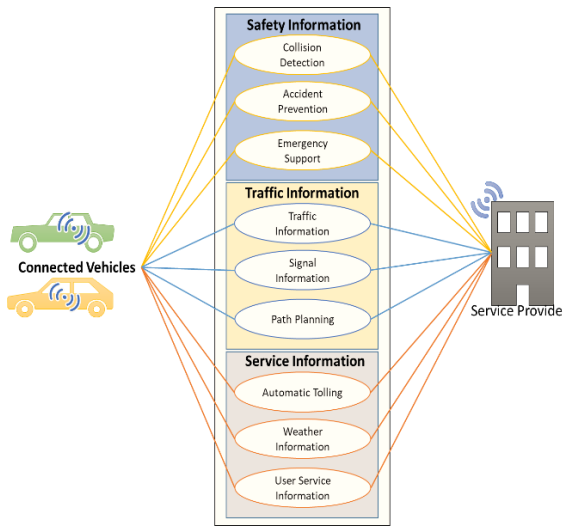

Fig.5. Use case diagram secure engineering model.

Data management system: System manages all information related on smart highway such as smart-tolling, transfer information that make a more efficient system to driver

A. Use case diagram of Engineering Modeling in ITS

Grasp related behavior between the driver and the smart highway group, we will find detailed in following Use case descriptions about the action run which order and what's related towards to each behavior. Based on these, we expect to get final goal of analysis the behavior and purpose of each sub-goal on goal-based approach. The table 1 has shown description of USE case of secure engineering design for connected vehicles.

TABLE 1. USECASE DESCRIPTION OF SECURE ENGINEEIRNG

| Use Case | Name | Pre-Condition | Post-Condition |
|---|---|---|---|
| UC1. | Smart-tolling | Connected successfully between mobile device in vehicle and main system | Calculate smart highway toll by minimizing decrease of vehicle speed |
| UC2. | Preventing of car accident | Network between Terminal Application and Command Center Application always should be maintained All of users are must enrolled in Smart Highway's member | Prevention of car accident's function provide intelligence information which can prevent car accidentby integrating terminal application's data and sensor's data using in smart highway |
| UC3. | Supporting information for emergency and disaster situation | Connected successfully between mobile device in vehicle and main system | Improve efficiency by checking information that is identified by sensor, and it builds a database |
| UC4 | Weather forecast | Connected successfully between mobile device in vehicle and main system | To prevent accidents due to bad weather |
| UC5 | Urgent Emergency notify | The emergency vehicle enter the smart highway, and main system detect the entry of vehicle | Handle the emergency successfully, so decrease threat of smart highway |
| UC6 | Using additional service in smart highway | Network between Terminal Application and Command Center Application always should be maintained | Actor's convenience and satisfaction arise by using additional service in Smart Highway |

## 5. IMPACT AND FINAL REMARKS

In upcoming years' transportations system will be completely modify. It will be more intelligent and autonomous vehicle around us such as smart car, auto driver, etc. Some of automotive industries already produce smart car. But our main concern about infrastructure of transportation system which support intelligent transportation. Current infrastructure will be not support intelligent transportation system. So, in this article, we are going to introduce secure intelligent transportation (roadside) infrastructure of user's (Driver, Autonomous driver etc.) point of view. It's called intelligent transportation infrastructure. The impact of connected vehicles in ITS will proved to be a break through on the existing infrastructure of transportation system. These are some of the following effects

*A. Social Impact*
- Provide better and effective transportation life.
- Provide safe and rapid transit information service under high-speed driving conditions.
- Satisfy the driver's driving service and improve

the quality of life
- Provide traffic information service that fuses with telematics
- Provide continuous ITS service in time and space through expansion and linkage of existing ITS technologies

*B. Business Impact*

- Core technology pre-emption related to ITS service as next-generation growth power and synergy effect of technical
- By exporting developed technology, be more competitive in the world market and exploit a way out of export with all technology that need to build test bed
- Expect creating a ripple effect of related industry: have direct and indirect influence on widespread industry such as vehicle, wireless communication, mobile communication terminal, internet, m- commerce etc.
- Promote the national economy by reducing personnel and materiel cost in traffic congestion and traffic accidents that may occur in road under the goal of ITS technical development of 'accident- free' and 'nonstop'
- Inducement the creation of new employment about spreading industry related to ITS: by invigorating the ITS services, promote domestic industry and induce the creation of new employment

*C. Technical Impact*

- In high-speed driving environment, guiding role of technology related to real-time DB processing technology development and middleware technology development
- Guarantee of technology caused by alliance of telematics services as part of the national policy and Intelligent Roadside infrastructure.
- Globally source technology secure caused by developing the domestic technology by designing the architecture related to developing a traffic monitoring and traffic information fusion technology under the high-speed traffic condition the domestic technology

However, our aim to provide safe and secure infrastructure for smart vehicles (autonomous vehicles) etc.

**ACKNOWLEDGEMENT**

This work was supported by Institute for Information & communications Technology Promotion (IITP) grant funded by the Korea government (MSIT) (No.2017-0-00560, Development of a Blockchain based Secure Decentralized Trust network for intelligent vehicles).